\documentclass[12pt]{article}

\usepackage{graphicx}

\def\rf#1{(\ref{eq:#1})}
\def\lab#1{\label{eq:#1}}

\def\br{\begin{eqnarray}}
\def\er{\end{eqnarray}}
\def\be{\begin{equation}}
\def\ee{\end{equation}}
\def\({\left(}
\def\){\right)}
\def\pa{\partial}
\def\u2{\mid u\mid^2}
\def\rlx{\relax\leavevmode}
\def\IR{\rlx\hbox{\rm I\kern-.18em R}}

\begin{document}

\begin{titlepage}
\vspace*{-1cm}

\vskip 3cm

\vspace{.2in}
\begin{center}
{\large\bf A model for Hopfions on the space-time $S^3\times \IR$}
\end{center}

\vspace{.5cm}

\begin{center}
E. De Carli~$^{2,3}$ and L. A. Ferreira~$^{1,2}$

\vspace{.5 in}
\small

\par \vskip .2in \noindent
$^{(1)}$Instituto de F\'\i sica de S\~ao Carlos; IFSC/USP;\\
Universidade de S\~ao Paulo  \\ 
Caixa Postal 369, CEP 13560-970, S\~ao Carlos-SP, Brazil\\

\par \vskip .2in \noindent
$^{(2)}$~Instituto de F\'\i sica Te\'orica - IFT/UNESP\\
Universidade Estadual Paulista\\
Rua Pamplona 145\\
01405-900,  S\~ao Paulo-SP, Brazil\\

\par \vskip .2in \noindent
$^{(3)}$~Departamento de Engenharia Mec\^anica - LMPT\\
Universidade Federal de Santa Catarina - UFSC\\
Campus Universit\'ario - Trindade \\
88040-900, Florian\'opolis-SC, Brazil

\normalsize
\end{center}

\vspace{.5in}

\begin{abstract}
We construct static and time dependent exact  soliton solutions for a 
theory of scalar fields taking values on a wide class of two dimensional target
spaces, and defined  on the four dimensional space-time $S^3\times 
\IR$. The construction is based on an ansatz built out of special 
coordinates on $S^3$. The requirement for finite energy introduces 
boundary conditions that determine an infinite discrete spectrum of
frequencies for the oscillating solutions. For the case
where the target space is the sphere $S^2$, we obtain static soliton
solutions with non-trivial Hopf topological charges. In addition, such
hopfions can oscillate in time, preserving their topological
Hopf charge, with any of the frequencies belonging to that
infinite discrete spectrum.   

\end{abstract} 
\end{titlepage}

\section{Introduction}

In this paper we consider a non-linear theory of a complex scalar
field $u$, on the space-time
$S^3\times \IR$.  Our considerations apply to a wide class of target
spaces, but the case of most interest is that of the sphere $S^2$, in
which case the field $u$ parametrizes a plane corresponding to the
stereographic projection of $S^2$. 
The action is the integral of the square of the
pull-back of the area form on the target space. Therefore, it is
quartic in derivatives, but only quadratic in time derivatives.
The theory is integrable in the sense that it possesses a
generalized zero curvature representation of its equations of motion
 and an infinite number of local conservation laws
 \cite{afs97,bf}. The conserved currents are associated to the
 invariance of the theory under the area preserving diffeomorphisms on the
 target space.  

We construct an infinite number of static and time dependent exact
soliton solutions using an 
ansatz \cite{afz99,bf} that reduces the four dimensional non-linear
equations of motion into  linear ordinary differential
equations for the profile function. In the case of the target space
$S^2$ the solitons carry non-trivial Hopf topological charges.
Although the topology is
the same as that of other models 
possessing hopfion solutions \cite{afz99,sf,nicole}, the Derrick's
scaling arguments is circumvented in a different manner.
The stability of the static three dimensional
solutions comes from the fact that the physical space is $S^3$ and
that introduces a length scale given by its radius. A model in
Euclidean space where a similar stability 
mechanism occurs is discussed in \cite{laf}.  

The requirement for finite energy leads to boundary
conditions that determine an infinite discrete set of allowed
frequencies for the oscillations of the solutions. Those solutions 
can be linearly superposed since the profile function  satisfies a
linear equation. It turns out that the energy of the superposed
solution is the sum of the energies of the modes, and in this sense the modes 
are decoupled. However, the profile function can take values only on some
intervals on the real line, which depend on the choice of the target
space. Therefore, not all superpositions are allowed and that
introduces some sort of coupling among the modes. The allowed
superpositions for the case of the target space $S^2$ are discussed in
detail. One of the most interesting superpositions corresponds to
oscillating hopfion solutions. We show that it is possible to superpose
to the static hopfion soliton any number of oscillating modes with any
of the frequencies belonging to that infinite discrete
spectrum. Although the solution oscillates in time, its topological
Hopf charge is preserved. The only constraint on such superposition appears on
the intensity of each mode.

The paper is organized as follows: in section \ref{sec:model} we
introduce the model and discuss its integrability properties, in section
\ref{sec:ansatz} we propose the ansatz and construct the exact soliton
solutions, in section \ref{sec:energy} we discuss the energy, boundary
conditions and allowed frequencies. The case of the target space $S^2$
is discussed in section \ref{sec:s2}.

\section{The model}
\label{sec:model}

The metric on the space-time $S^3\times \IR$ is given by 
\be
ds^2=  dt^2 - r_0^2\, \( \frac{dz^2}{4\, z\(1-z\)}+\(1-z\) d{\varphi_1}^2+ 
z\, d{\varphi_2}^2 \)
\lab{metric}
\ee
where $t$ is the time,  $z$, and $\varphi_i$,
$i=1,2$, are coordinates on the sphere $S^3$, and $0\leq z \leq 1$,
$0\leq \varphi_i \leq 2 \pi$, and $r_0$ is the radius of the sphere
$S^3$. Embedding $S^3$ on $\IR^4$ we get that
the Cartesian coordinates of the points of $S^3$ are \br
x_1&=& r_0\, \sqrt{z}\, \cos \varphi_2 \qquad \qquad 
x_3= r_0\,\sqrt{1-z}\, \cos \varphi_1 \nonumber\\  
x_2&=& r_0\,  \sqrt{z}\, \sin \varphi_2 \qquad \qquad \,
x_4=  r_0\,\sqrt{1-z}\, \sin \varphi_1
\lab{cartcoords3}
\er 
The model is defined by the action 
\be
S= \int dt\; \int_{S^3} d\Sigma  \; \frac{h^2_{\mu\nu}}{ \gamma^2}
\lab{action}
\ee
where the volume element on $S^3$ is $d\Sigma = \frac{r_0^3}{2}\, dz \,
d\varphi_1\, d\varphi_2$, and 
\be
h_{\mu\nu}\equiv \partial_{\mu} u\partial_{\nu} u^* - 
 \partial_{\nu} u \partial_{\mu} u^*
\lab{hdef}
\ee
where $\pa_{\mu}$ denotes partial derivatives with respect to the four 
coordinates on $S^3\times \IR$, namely $\zeta^{\mu}\equiv \(t, z,
\varphi_1,\varphi_2\)$, $u$ is  a
complex scalar field, and $\gamma \equiv \gamma 
\(\mid u \mid^2\)$, is a real functional of the squared modulus of $u$, and
it defines the geometry of the target space.  The metric on target
space is given by
\be
d \sigma^2 = \frac{d u \, du^*}{\gamma}
\lab{targetmetric}
\ee
Some cases of
interest are the following: {\em a)} $\gamma = 1$ corresponding to the plane
with coordinates being the real and 
imaginary parts of $u$, {\em b)} $\gamma = \(1-\u2\)^2$, with $\mid u
\mid^2<1$,  corresponding
to the Poincar\'e hyperbolic disc, and {\em c)} $\gamma = \(1+\u2\)^2$
corresponding to the sphere $S^2$. In such case, $u$ is related to the
three dimensional unit vector ${\vec n}$ (${\vec n}^2=1$) defining
the sphere, through 
the stereographic projection 
\be
{\vec n} = \frac{1}{1+\u2}\,\(u+u^*,-i\(u-u^*\),\u2 -1\)
\lab{ndef}
\ee
Notice that in the case of the target space being $S^2$, the action
\rf{action} corresponds to the quartic term of the Skyrme-Faddeev
action \cite{sf} which has been studied on the space-time $S^3\times
\IR$ in \cite{ward}.

The Euler-Lagrange equations following from \rf{action} are given by 
\be
\partial^{\mu} \(\frac{K_{\mu}}{\gamma}\)=0
\lab{eqmot}
\ee
together with its complex conjugate, and where 
\be
K_{\mu} = h_{\mu\nu} \pa^{\nu} u = \( \pa u \, \pa u^*\) \, \pa_{\mu}
u - \(\pa u\)^2 \, \pa_{\mu} u^*
\lab{kdef}
\ee

The model \rf{action} possesses an infinite
number of local conserved currents given by
\be
J_{\mu} =  \frac{K_{\mu}}{\gamma}\, \frac{\delta G}{\delta u} - 
\frac{K_{\mu}^*}{\gamma}\, \frac{\delta G}{\delta u^*}
\lab{conservcur}
\ee
where $G$ is a function of $u$ and $u^*$ but not of its derivatives. 
Using the equations of motion \rf{eqmot} and the identities 
\be
K_{\mu} \, \pa^{\mu} u=0  \qquad \qquad 
K_{\mu} \, \pa^{\mu} u^* - K_{\mu}^* \, \pa^{\mu} u =0
\ee
one can check that \rf{conservcur} is indeed conserved,
i.e. $\pa^{\mu}J_{\mu}=0$. The symmetries associated to such
conservation laws are the area preserving diffeomorphisms of the
target manifold. Indeed, the tensor $h_{\mu\nu}/\gamma$ is the
pull-back of the area form on the target space
\be
dA = i \frac{du\wedge du^*}{\gamma}
\lab{areaform}
\ee
Therefore, the action \rf{action} is invariant under 
diffeomorphisms  preserving the area \rf{areaform}, and
\rf{conservcur} are the corresponding Noether currents \cite{fr,bf}. 

\section{The ansatz}
\label{sec:ansatz}

Following \cite{bf,afz99} we introduce the ansatz 
\be
u = f\(t,z\) \, e^{i\( m_1 \varphi_1 + m_2 \varphi_2\)}
\lab{ansatz}
\ee
where $m_i$, $i=1,2$, are arbitrary integers, and $f$
is a real profile function. Replacing it into the equation of motion
\rf{eqmot} one gets 
\be
\partial_t \( \frac{\partial_t f^2}{\gamma}\) - 
\frac{4\,z\,\(1-z\)}{r_0^2\,\Omega}\, \partial_z\(\Omega\, \frac{\partial_z
  f^2}{\gamma}\) = 0
\lab{feq}
\ee
where
\be
\Omega \equiv m_1^2 z + m_2^2 \(1-z\)
\lab{omegadef}
\ee
We now make a change of variable in the profile function, introducing 
a function $g$ by
\be
d g \sim \frac{d f^2}{\gamma} 
\lab{changef}
\ee
For instance, in the particular cases discussed below \rf{targetmetric} we get
that: 
{\em a)} for the plane where $\gamma =1$ one has $g=f^2$, and $g \geq
0$;  {\em b)} for the Poincar\'e hyperbolic disk where $\gamma =
\(1-\u2\)^2$, one has $g= 1/\(1-f^2\)$, and $g\geq 1$;   
{\em c)} for the sphere
$S^2$ where $\gamma = \(1+\u2\)^2$ one has $g= 1/\(1+f^2\)$, and  $0\leq g
  \leq 1$.

So, with the change \rf{changef} one gets that \rf{feq} becomes a
linear partial differential equation in $g$
\be
\partial_t^2 g -  
\frac{4\,z\,\(1-z\)}{r_0^2\,\Omega}\, \partial_z\(\Omega\,\partial_z g\) = 0
\lab{pdeg}
\ee
The solutions can be obtained by separation of
variables introducing 
\be
g\(t,z\)= J\(t\) \, H\(z\)
\lab{separablesol}
\ee
and then $J$ and $H$ have to satisfy 
\be
\pa_t^2 \, J + \omega^2 \,J=0
\lab{jeq}
\ee
and
\be
z\(1-z\)\, \pa_z\left[ \(q^2 z + \(1-z\)\) \, \pa_z \, H\right] +
\frac{r_0^2\,\omega^2}{4}\, 
\(q^2 z + \(1-z\)\) \, H =0
\lab{heq}
\ee
where $\omega^2$ is the separation of variables constant, and where $q$ is
\be
q\equiv \frac{\mid m_1\mid}{\mid m_2\mid}
\lab{qdef}
\ee

The equation \rf{heq} is what is called a Heun equation
\cite{heun}. It is a generalization of Gauss hypergeometric equation
in the sense that it has one extra regular singular point. Inddeed, 
\rf{heq} reduces to the hypergeometric equation for $q=1$. Notice that
\rf{heq} is invariant under the joint transformations $q
\leftrightarrow 1/q$ and $z \leftrightarrow 1-z$. Therefore, if
$H_{q,\omega}\(z\)$ is a solution of \rf{heq} for some value of $q$, so is
\be
H_{{1/q,\omega}}\(z\)=H_{q,\omega}\(1-z\)
\lab{invrel}
\ee
for the inverse value $1/q$. One can  obtain solutions 
of \rf{heq}  in powers series around $z=0$ and for $0\leq q\leq 1$.
The solutions for $q\geq 1$, are then obtained from those 
using the above symmetry. The power series solutions are given by
\be
H_{q,\omega}\(z\)= \frac{1}{v_{q,\omega}}\;\( z + \sum_{n=2}^{\infty}
c_n z^n\) 
\lab{hqnorm}
\ee
where the positive constants $v_{q,\omega}$ are chosen such that the
maximum absolute value of $H_{q,\omega}\(z\)$ in the interval
  $0\leq z \leq 1$, is unity. The coefficients, for $n\geq 2$, are
determined by the recursion relation (with $c_0=0$, and $c_1=1$) 
\br
&c_n& = \frac{-1}{n\(n-1\)}\left[ 
\(q^2-1\)\(\frac{\omega^2\, r_0^2}{4} -
\(n-2\)^2\)\, c_{n-2}
\right. \lab{recursion}\\
&+& \left. 
\(\frac{\omega^2\, r_0^2}{4}-\(n-1\)\(n-2\) 
+ \(q^2-1\) \(n-1\)^2\)\, c_{n-1}
\right] \nonumber
\er
We will be interested in solutions with $\omega$ real, and therefore
the solutions of \rf{jeq} are trigonometric functions. 
Of course, if $g$ is a solution of \rf{pdeg}, so is $\alpha \, g +
\beta$, with $\alpha$ and $\beta$ constants. We will then use  the
following normalization for the solutions of \rf{pdeg}
\be
g_{q,\omega}= \frac{1}{2}\left[
\sin\(\omega \, t +\delta\) \, H_{q,\omega}\(z\) + 1\right] \qquad \qquad 
\omega \neq 0
\lab{normtimesol}
\ee
with $H_{q,\omega}\(z\)$ given by \rf{hqnorm}. The advantage of such
normalization is that the solutions \rf{normtimesol} take values on
the real line 
from zero to unity  only. Therefore, they are admissible solutions for
the case where the target space is the sphere $S^2$ (see discussion
below \rf{changef}). 

We normalize the static solutions of \rf{pdeg} as 
\be
g_{q,0} = \frac{\ln\( q^2 \, z + 1 - z\)}{\ln q^2}  
\lab{solgs2}
\ee
So, $g_{q,0}$ is a monotonic function varying
from zero at $z=0$ to unity at $z=1$. Notice that $g_{q,0} \rightarrow 
z $ as $q \rightarrow 1$. A decreasing function can be
obtained by the interchange $g_{q,0}\rightarrow 1-g_{q,0}$. Therefore,
\rf{solgs2} 
are also admissible solutions for the case where the target space is
the sphere $S^2$. 

The admissible solutions for the cases of the plane or the  Poincar\'e
hyperbolic disk can then be written as 
\be
g_{q,\omega}^{\rm plane} = \alpha \, g_{q,\omega}
\qquad \qquad 
g_{q,\omega}^{\rm Poinc. disk} = \alpha \,  g_{q,\omega} + 1
\ee
with $\alpha$ being a real and positive constant, and $g_{q,\omega}$
being given either by \rf{normtimesol} or \rf{solgs2}. 

\section{The energy and boundary conditions}
\label{sec:energy}

The energy for the solutions obtained
through the ansatz \rf{ansatz}, \rf{changef}, is given by
\be
E =  \int_{S^3} d\Sigma  \; {\cal H} = 
16\,\pi^2\, r_0\,\; \int_0^1 dz\; \Omega\, \( \frac{\(\partial_t
  g\)^2}{4\,z\,\(1-z\)} + \frac{1}{r_0^2}\;\(\pa_z g\)^2\)
\lab{energy}
\ee
where ${\cal H}$ is the Hamiltonian density associated to \rf{action}. 
Using the equation of motion \rf{pdeg} one gets that the energy of
static configurations is
\be
E_{\rm static} = \frac{16\,\pi^2}{r_0}\; \( \Omega \, g \, \partial_z
g\)\mid_{z=0}^{z=1} 
\ee
Therefore, the energy of the static solutions \rf{solgs2} is 
\be
E\( g_{q,0}\)= \frac{16\,\pi^2}{r_0}\; \mid m_1\, m_2\mid \;
\frac{\(q-1/q\)}{\ln q^2}
\lab{energystatics2sol}
\ee

Notice
that the solutions of \rf{heq}, for $\omega\neq 0$, which do not
vanish at $z=0$ or $z=1$, have a logarithmic divergence on its
first derivative, at those points. We then observe from
\rf{energy} that such solutions do not have finite
energy. Consequently, we shall impose the following boundary
conditions
\be
H_{q,\omega}\(0\) = H_{q,\omega}\(1\) =0 \qquad \qquad {\rm for} \quad
\omega\neq 0
\lab{boundarycond}
\ee  
In addition, using the equation of motion \rf{pdeg} one obtains that
\be
\frac{d\, E}{d\,t} = \frac{32 \, \pi^2}{r_0} \,\( \Omega\, \partial_t g \,
\partial_z g \)\mid_{z=0}^{z=1}
\ee
Therefore, the energy is conserved for the static solutions, and for
those solutions \rf{normtimesol} satisfying 
the boundary conditions \rf{boundarycond}. 

Multiplying \rf{heq} by $H_{q,\omega}$ and integrating in $z$, one gets
that
\be
\int_0^1 dz \, 
\frac{\omega^2\,\Omega}{4z\(1-z\)}\, H_{q,\omega}^2 -
\frac{1}{r_0^2}\, \int_0^1 dz \,\Omega\, \(H_{q,\omega}^{\prime}\)^2 = - 
\frac{1}{r_0^2}\,\(\Omega\,H_{q,\omega}\,H_{q,\omega}^{\prime}
\)\mid_{z=0}^{z=1} 
\ee 
Consequently, the energy \rf{energy} for the solutions
\rf{normtimesol} satisfying \rf{boundarycond} is
\be
E\(g_{q,\omega}\)= \frac{16 \,\pi^2}{r_0} \, \mid m_1 \, m_2\mid \; 
\Lambda_{q,\omega}
\lab{energytimesol}
\ee  
where
\be
\Lambda_{q,\omega} = \Lambda_{1/q,\omega} = 
\frac{\omega^2\, r_0^2}{16}\,\int_0^1 dz \, 
 \( \frac{q}{1-z}  + \frac{1}{q\, z}\)\,
H_{q,\omega}^2=\frac{1}{4}\,
\int_0^1 dz \,\( q \, z + \frac{1-z}{q}\)\,
\(H_{q,\omega}^{\prime}\)^2
\lab{lambdaqomega}
\ee
The fact that $\Lambda_{q,\omega}$ is invariant under
$q\leftrightarrow 1/q$, follows from the symmetry \rf{invrel}. In
Table \ref{tab:lambdaqomega} we give the values of $\Lambda_{q,\omega}$
for some chosen values of $q$ and some allowed frequencies $\omega$.
So, the energies \rf{energystatics2sol} and \rf{energytimesol} do
not depend upon the signs of the integers $m_1$ 
and $m_2$ and it is invariant under the interchange $m_1
\leftrightarrow m_2$ (see \rf{qdef}).

Multiplying \rf{heq} by $H_{q,{\bar\omega}}$, subtracting from the
same relation with $\omega$ interchanged with ${\bar\omega}$, and
integrating in $z$, one gets that
\be
\(\omega^2-{\bar \omega}^2\)\, \int_0^1 dz \, 
\frac{\Omega}{4z\(1-z\)}\, H_{q,\omega}\,H_{q,{\bar\omega}} = 
\frac{1}{r_0^2}\, 
\(\Omega\, \( H_{q,\omega}\,H_{q,{\bar\omega}}^{\prime} -
 H_{q,\omega}^{\prime}\,H_{q,{\bar\omega}}\)\)\mid_{z=0}^{z=1}
\ee
Similarly, differentiating \rf{heq} w.r.t. $z$ once, multiplying by
$H_{q,{\bar\omega}}^{\prime}$, subtracting from the same relation with
$\omega$ interchanged with ${\bar\omega}$, and 
integrating in $z$, one gets that
\be
\(\omega^2-{\bar \omega}^2\)\, \int_0^1 dz \, 
\Omega\, H_{q,\omega}^{\prime}\,H_{q,{\bar\omega}}^{\prime} = 
\frac{1}{r_0^2}\, 
\(4z\(1-z\)\,\Omega\, 
\( H_{q,\omega}^{\prime}\,H_{q,{\bar\omega}}^{\prime\prime} -
 H_{q,\omega}^{\prime\prime}\,H_{q,{\bar\omega}}^{\prime}\)\)\mid_{z=0}^{z=1}
\ee
Consequently, for the solutions \rf{normtimesol} satisfying
\rf{boundarycond}, and having finite first and second $z$-derivatives
at $z=0$ and $z=1$, one gets the orthogonality relations 
\be
\int_0^1 dz \, 
\frac{\Omega}{4z\(1-z\)}\, H_{q,\omega}\,H_{q,{\bar\omega}} =
\int_0^1 dz \, 
\Omega\, H_{q,\omega}^{\prime}\,H_{q,{\bar\omega}}^{\prime} = 0 
\qquad \qquad {\rm for} \quad \omega
\neq {\bar\omega}
\ee
It then follows that the energy \rf{energy} of a linear combination of
solutions \rf{normtimesol} satisfying
\rf{boundarycond}, is just the sum of the energies of each solution, i.e. 
\be
E\(A\, g_{q,\omega}+ B\, g_{q,{\bar\omega}}\) = A^2 E\( g_{q,\omega}\)
+ B^2 E\(g_{q,{\bar\omega}}\) \qquad \qquad {\rm for} \quad \omega
\neq {\bar\omega}
\lab{decompenergy}
\ee
Therefore, in that sense, the modes for the same value of $q$ are
decoupled. However, the intensities in which they enter in the
superposition are not independent. As discussed below
\rf{changef} the real values that the profile function $g$ can take 
depend on the target space under consideration. 
So, when we take linear combinations of the solutions we have to
respect those constraints. In section \ref{sec:s2} we  discuss those
constrained linear combinations in detail in the case
where the target space is the sphere $S^2$. 

The boundary conditions \rf{boundarycond} lead to a discrete spectrum
of allowed frequencies $\omega$. Indeed, for the case where $q=1$ the series
\rf{hqnorm} truncates whenever
\be
\frac{r_0^2\, \omega^2}{4} = n\(n+1\) \qquad \qquad \qquad
n=1,2,3,\ldots
\lab{discreteomegaq1}
\ee
and the corresponding polynomials satisfy the boundary conditions
\rf{boundarycond}. The first four of those polynomials are given by
(with the normalization as in \rf{hqnorm})
\br
H_{1,2}&=& 4\( z-z^2\) \nonumber\\
H_{1,6}&=& 6\sqrt{3}\( z-3 z^2 + 2 z^3\) \lab{polyq=1}\\
H_{1,12}&=& 16\( z - 6 z^2  + 10 z^3  - 5 z^4\) \nonumber\\
H_{1,20}&=& \frac{2450\,\(z - 10 z^2  + 30 z^3  - 35 z^4  + 14 z^5\)}{
\sqrt{7}\;(3\sqrt{5} + 5\sqrt{6})\sqrt{15 - 2\sqrt{30}}}
 \nonumber
\er
where the first index refers to $q=1$ and the second to $n\(n+1\)$ as
given by \rf{discreteomegaq1}.  

For $q\neq 1$ the series \rf{hqnorm} does not truncate, and the
frequencies $\omega$ leading to solutions satisfying \rf{boundarycond}
can be found numerically. We give in Table \ref{frequencies} the first
three frequencies for some chosen values of $q$.  In addition, the
same frequencies hold true 
under the exchange $q\leftrightarrow 1/q$ due to the symmetry
\rf{invrel}. Therefore, the frequencies become smaller as $q$ departs
from unity, either to smaller or greater values than unity.

\begin{figure}[h]
\centering 
	\includegraphics[scale=1.0]{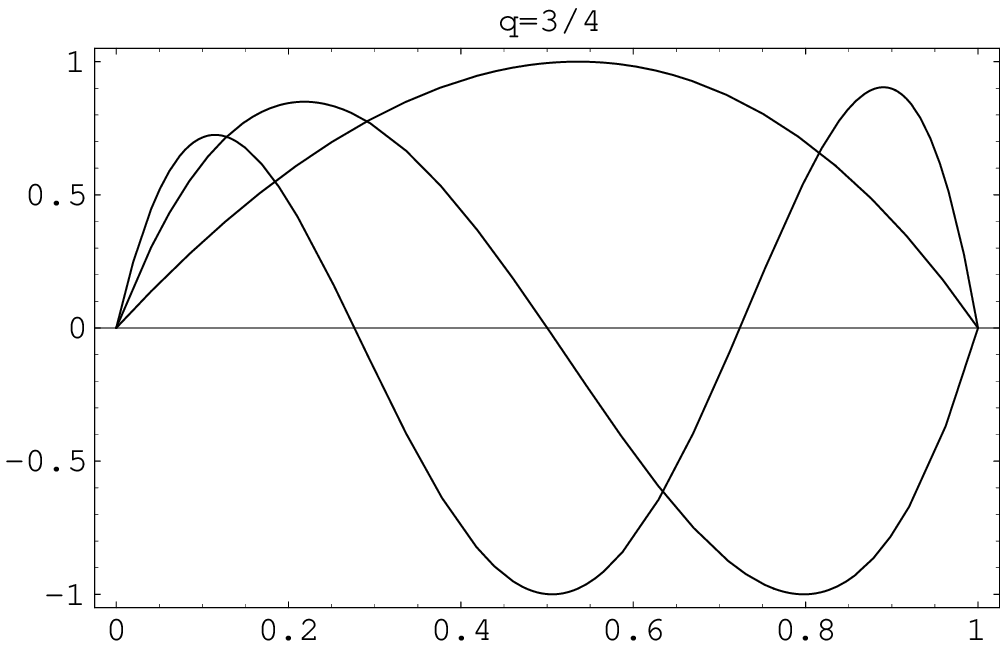}
\vspace{1 cm}
        \includegraphics[scale=1.0]{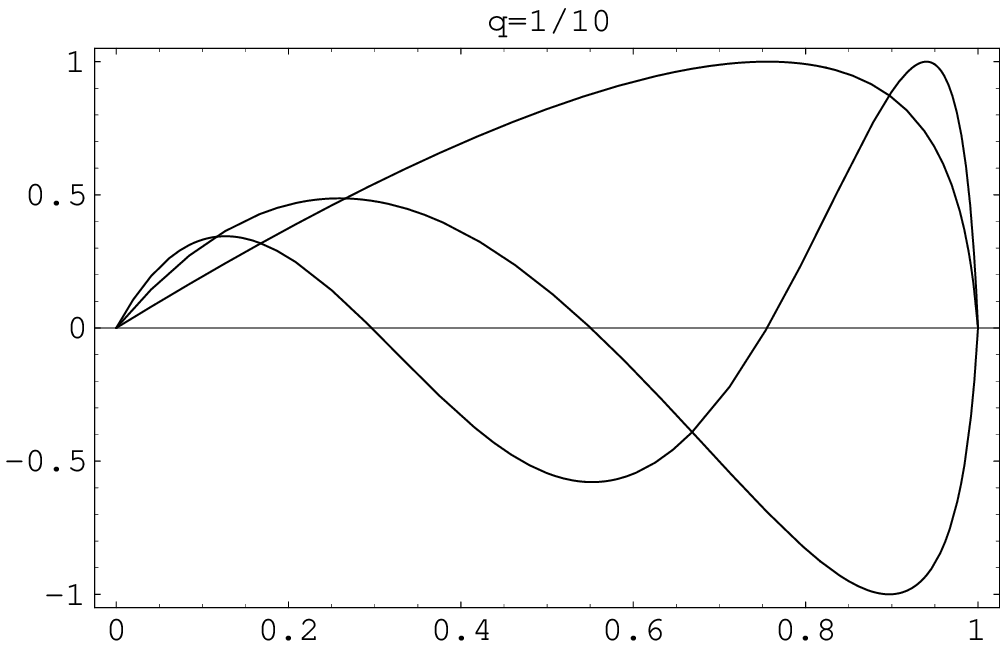}
	\caption[modes]{\parbox[t]{.7\textwidth}{Plots of
	$H_{q,\omega_i}\( z\)$, $i=1,2,3$, as normalized in \rf{hqnorm},
	for $q=3/4$ (top) and $q=1/10$ (bottom) and  for the three
	frequencies given in Table 
	\ref{frequencies}. The number of zeroes increase with the
	increase of $\omega_i$. The plots of $H_{q,\omega}$ and
	$H_{1/q,\omega}$ are related by reflection around $z=1/2$ due
	to the symmetry \rf{invrel}.}}  
	\protect\label{fig:modes}
\end{figure}

In Figure \ref{fig:modes} we exemplify the shape of the functions
$H_{q,\omega}$, by plotting  the first three modes
$H_{q,\omega_i}$, for $q=3/4$ and $q=1/10$ and the frequencies $\omega_i$,
$i=1,2,3$, given in  Table \ref{frequencies}. Due to the symmetry
\rf{invrel}, $H_{1/q,\omega_i}$ can be obtained by reflecting the
plots 
around $z=1/2$. Notice that the polynomial solutions for $q=1$, 
given in \rf{polyq=1},  are invariant under \rf{invrel}.  One
observes that as $q$ decreases, the functions 
$H_{q,\omega}$ get deformed in a way that their first derivatives at $z=1$
increase. For $q>1$, it follows, due to \rf{invrel},  that the first
derivative of $H_{q,\omega}$ at $z=0$ increases as $q$ increases.

\section{The case of $S^2$ as target space}
\label{sec:s2}
 
According to the comments below \rf{changef}, 
in the case  where the target space is the two dimensional
sphere $S^2$, we have that $\gamma = \(1+\u2\)^2$, and so $g=1/\(1+f^2\)$ and
$0\leq g\leq 1$. The ansatz \rf{ansatz} becomes
\be
u = \sqrt{\frac{1-g}{g}}\, e^{i\( m_1 \varphi_1 + m_2 \varphi_2\)}
\lab{ansatzs2}
\ee

As we have discussed, the solutions \rf{normtimesol} and \rf{solgs2}
are admissible solutions for the case where the target space is
$S^2$. However, one can construct more admissible solutions, i.e. with
$0\leq g\leq 1$, by taking linear combinations of the solutions
\rf{normtimesol} and \rf{solgs2}. An interesting case corresponds to
linear combinations of the static solution $g_{q,0}$ given by
\rf{solgs2}, with one or 
more time dependent solutions of the type \rf{normtimesol}. It leads
to oscillating (in time) hopfion solutions. Let us consider the case
$q\leq 1$ first. The static solution $g_{q,0}$ vanishes at $z=0$, and
has $z$-derivative 
 equals to $\(q^2-1\)/\ln q^2$ there. The function $f_q\(z\)\equiv 
\left[\(q^2-1\)/\ln q^2\right]\, v_{q,\omega} \, H_{q,\omega}\(z\)$,
with $H_{q,\omega}$ given by \rf{hqnorm} and satisfying
\rf{boundarycond}, has the same behaviour at $z=0$ as $g_{q,0}$.  
In addition, $c_2$ given by \rf{recursion}, is negative for the case
of $H_{q,\omega}$ satifying \rf{boundarycond}, and so $f_q\(z\)$ grows
slower than  $g_{q,0}$ for small $z$. 
We do
not have a rigorous proof, but by careful direct inspection we found
that,  as $z$ increases
from zero to unity the absolute value of $f_q\(z\)$ never exceeds the value
of $g_{q,0}$, which is a positive monotonic function of $z$. It then
follows that  the combination  
\be
g_{q,0} + \alpha \frac{\(q^2-1\)}{\ln q^2}\, v_{q,\omega} \,
\sin\(\omega t + \delta\)\, H_{q,\omega}\(z\)
\ee
with $0\leq \alpha \leq 1$, takes values between zero and unity only,
and so it is an admissible time dependent solution for the target
space $S^2$. Therefore, by adding up such of type of solutions and
dividing by their number, one gets that the time dependent solutions 
\be
g_q^{(N)}\(t,z\)= \frac{\ln\( q^2 \, z + 1 - z\)}{\ln q^2} + \frac{1}{N} 
\frac{\(q^2-1\)}{\ln q^2} \sum_{\omega_i} \alpha_i \, v_{q,\omega_i}\,
\sin\(\omega_i t +\delta_i\) H_{q,\omega_i}\(z\)
\lab{oscillatehopfion}
\ee
where $N$ is the number of modes (frequencies) entering into the sum,
and $\alpha_i$'s are  real coefficients satisfying $0\leq  \alpha_i
\leq 1$, are admissible solutions for the target space $S^2$. 
The energy of such solutions, according to \rf{normtimesol},
\rf{solgs2}, \rf{energystatics2sol}, 
\rf{energytimesol} and \rf{decompenergy}, is given by
\be
E = \frac{16\,\pi^2}{r_0}\; \mid m_1\, m_2\mid \;\frac{\(q-1/q\)}{\ln q^2}
\( 1 + 4 \, \frac{q^2}{N^2}\; \frac{\(q-1/q\)}{\ln q^2} \,  
\sum_{\omega_i} \alpha_i^2\, v_{q,\omega_i}^2\,\Lambda_{q,\omega_i} \)
\lab{oscillateenergy}
\ee
with $\Lambda_{q,\omega_i}$ given by \rf{lambdaqomega}. 

Solutions of the type \rf{oscillatehopfion} for $q\geq 1$ can be
obtained using the symmetry \rf{invrel}, i.e. $g_{1/q}^{(N)}\(t,z\)= 
g_q^{(N)}\(t,1-z\)$. Solutions that decrease from unity at $z=0$ to
zero at $z=1$ can be obtained by the symmetry $g_q^{(N)}\(t,z\) \rightarrow
1-g_q^{(N)}\(t,z\)$. The energy \rf{oscillateenergy} is invariant
under those two type of transformations. As we now show, all these
solutions correspond to oscillating 
hopfions, i.e. solutions that oscillates in time and have a constant (in time)
non-trivial Hopf number.

For any fixed time $t$ our solutions define a map from the physical space
$S^3$ to the target space $S^2$, and so it is a Hopf map
\cite{bott}. We now show that the Hopf invariant (the linking number)
is independent of time for the admissible solutions we have constructed. 
In order to calculate de Hopf index of the solution we introduce 
\br
\Phi_1 &=& \sqrt{g} \; \cos \(m_2 \varphi_2 \) \qquad \qquad \;\;
\Phi_3 = \sqrt{1-g} \; \cos \(m_1 \varphi_1 \) \nonumber\\
\Phi_2 &=& -\sqrt{g} \; \sin \(m_2 \varphi_2 \) \qquad \qquad 
\Phi_4 = \sqrt{1-g} \; \sin \(m_1 \varphi_1 \) 
\lab{s3phidef}
\er
which defines another $3$-sphere $S^3_{\Phi}$, since
$\Phi_1^2+\Phi_2^2+\Phi_3^2+\Phi_4^2=1$. 
The field $u$ in \rf{ansatzs2} can be written as 
\be
u = \frac{\Phi_3 + i \Phi_4}{\Phi_1+i\Phi_2}
\lab{hopf}
\ee
Since $u$ parametrizes the sphere $S^2$ through the stereographic
projection \rf{ndef}, we have that \rf{hopf} gives the  map
$S^3_{\Phi}\rightarrow S^2$. So, the Hopf index is in fact evaluated
through the map $S^3\rightarrow S^3_{\Phi}\rightarrow S^2$, as we now
explain.  
We introduce the potential
\be
{\vec A} = \frac{i}{2}\( Z^{\dagger} {\vec \nabla} Z - {\vec
  \nabla}Z^{\dagger}\; Z \) 
\ee
where
\be
Z = \( 
\begin{array}{c}
Z_1 \\
Z_2
\end{array}\) \; ;\qquad \qquad\qquad\qquad Z_1 = \Phi_3 + i \Phi_4\; ;
\qquad\quad 
Z_2 = \Phi_1+i\Phi_2
\lab{zdef}
\ee
and the differential operator ${\vec \nabla}$ is the gradient on the
physical space  $S^3$. The Hopf index is defined by the integral \cite{bott}
\be
Q_H = \frac{1}{4\pi^2} \, \int d \Sigma \; {\vec A}\cdot {\rm
  curl}\,{\vec A}
\ee
where $d \Sigma$ is the volume element on the physical space $S^3$. 
Evaluating we get
\be
{\vec A} =   - m_1 \frac{\(1-g\)}{\sqrt{1-z}}\; {\hat
  {\bf e}}_{\varphi_1} + m_2 \frac{g}{\sqrt{z}}\; {\hat
  {\bf e}}_{\varphi_2} 
\ee
and
\be
{\rm curl}\,{\vec A} = 2 \; \pa_z g \;\( 
- m_2 \sqrt{1-z}\; {\hat {\bf e}}_{\varphi_1} 
+ m_1 \sqrt{z}\; {\hat {\bf e}}_{\varphi_2} \)
\ee
Consequently we get that
\be
Q_H= m_1 m_2 \, \left[ g\(t,z=1\)-g\(t,z=0\)\right]
\ee
Therefore, the solutions \rf{solgs2} have Hopf index 
\be
Q_H\(g_{q,0}\)=m_1 m_2
\ee
and so they are static hopfion soliton solutions. The time dependent
solutions \rf{normtimesol} are constant at $z=0$ and $z=1$ due to the
boundary condition \rf{boundarycond},
i.e. $g_{q\omega} \(t,0\)= g_{q\omega} \(t,1\)=1/2$. Therefore, they
carry no Hopf number
\be
Q_H\(g_{q,\omega}\)= 0 \qquad \qquad \qquad \omega\neq 0
\ee
The solutions \rf{oscillatehopfion} although time dependent, also have
contant values at $z=0$ and $z=1$, determined by their static
component. It then follows  that
\be
Q_H\(g_q^{(N)}\)=m_1 m_2
\ee
and so they do correspond to oscillating hopfion soliton solutions. 

The Hopf  index can also be calculated as the linking number of the
pre-images of two points of $S^2$ \cite{bott}. Notice, from \rf{ndef}, that the
north pole of $S^2$, ${\vec n} = \(0,0,1\)$, corresponds to
$u\rightarrow \infty$, and so from \rf{ansatzs2} to $g=0$. On the
other hand, the south pole of $S^2$, ${\vec n} = \(0,0,-1\)$, corresponds to
$u=0$, and so  to $g=1$. Therefore, from \rf{s3phidef} and \rf{zdef},
we see that the pre-image on $S^3_{\Phi}$ of the north pole of $S^2$
correponds to $Z_1=e^{im_1\varphi_1}$ and $Z_2=0$, whilst the pre-image
  of the south pole to $Z_1=0$ and $Z_2=e^{-im_2\varphi_2}$. For the
  static solution \rf{solgs2} and the time dependent solutions
  \rf{oscillatehopfion}, the pre-images in the spacial $S^3$ of these two
  circles in $S^3_{\Phi}$ are constant in time. In addition, those two
  circles in $S^3_{\Phi}$ pass through each other $m_1m_2$ times as
  $\varphi_1$ and $\varphi_2$ 
  varies from $0$ to $2\pi$ in $S^3$. So their linking number is $m_1m_2$, and
  that is the Hopf index. For the time dependent solutions
  \rf{normtimesol} it is not possible to have $g=0$ and $g=1$ on
  different points on the spatial $S^3$ at the same time
  $t$. Therefore, the pre-images of the north and south poles of $S^2$
  never link, and so the Hopf index of such solutions vanishes.

\vspace{1cm}

\noindent {\bf Acknowledgements}  We are greatful to Olivier Babelon
for many helpful discussions throughout the development of this
work. LAF thanks the hospitality at LPTHE-Paris where part of this
work was developed. The project was funded by a Capes/Cofecub
agreement. LAF is partially supported by a  CNPq research
grant and  EDC  by a Fapesp scholarship.

\newpage

\begin{table}[h]
\begin{tabular}{|c||l|l|l|}
\hline
$q$ & $\qquad\;\;\;\omega_1^2r_0^2/4 $ &
$\qquad\;\;\;\omega_2^2r_0^2/4 $ &   $\qquad\;\;\;\omega_3^2r_0^2/4 $ \\ 
\hline \hline
1 &    2  &                     6          &          12 \\
\hline
3/4 &  1.983765266031 &     5.987866563654  &   11.988820961965  \\
\hline
1/2 &  1.913643550920 &     5.923271548463   &  11.930495946441 \\
\hline
1/4 &  1.733644829967 &     5.672599570202   &  11.668111143086 \\
\hline
1/10 & 1.517649738276 &     5.238029617258   &  11.066608078277  \\
\hline
\end{tabular}
\caption{The first three frequencies leading to solutions \rf{hqnorm}
  satisfying the boundary conditions \rf{boundarycond} for some chosen
  values of $q$. Due to the symmetry \rf{invrel} the frequencies are 
  invariant under the interchange $q\leftrightarrow 1/q$. }
\label{frequencies}
\end{table}

\vspace{1cm}

\begin{table}[h]
\begin{tabular}{|c||r|r|r|}
\hline
$q$ & $\;\;\;\Lambda_{q,\omega_1} $ &
$\;\;\;\Lambda_{q,\omega_2} $ &   $\;\;\;\Lambda_{q,\omega_3}$ \\ 
\hline \hline
1     &   4/3       &            27/5       &     64/7  \\
\hline
3/4   &   1.362643  &   4.693750  &     9.501984\\
\hline
1/2   &   1.505658  &   4.275566  &     9.428440\\
\hline
1/4   &   2.065180  &   4.819658  &     8.855009\\
\hline
1/10  &   3.739188  &   7.930185  &    13.004162\\
\hline
\end{tabular}
\caption{Numerical values of $\Lambda_{q,\omega_i}$, as defined in
  \rf{lambdaqomega}, for some chosen
  values of $q$ and for the  three frequencies given in Table
  \ref{frequencies}. Due to \rf{invrel}, $\Lambda_{q,\omega}$ is invariant
  under the interchange $q \leftrightarrow 1/q$.} 
\label{tab:lambdaqomega}
\end{table}

\vspace{1cm}

\begin{table}[h]
\begin{tabular}{|c||r|r|r|}
\hline
$q$ & $\;\;\;v_{q,\omega_1} $ &
$\;\;\;v_{q,\omega_2} $ &   $\;\;\;v_{q,\omega_3}$ \\ 
\hline \hline
1     &    1/4        &  1/6$\sqrt{3}$  &   1/16    \\
\hline
3/4   &   0.28518993  & 0.11915683    &  0.07078933  \\
\hline
1/2   &   0.33261165 &  0.15277952  &    0.08701028 \\
\hline
1/4   &   0.40999103 &  0.20414992   &   0.12695732  \\
\hline
1/10  &   0.50274925  & 0.25672647  &    0.16731092 \\
\hline
\end{tabular}
\caption{Numerical values of the normalization constant
  $v_{q,\omega_i}$, as defined in \rf{hqnorm}, for some chosen
  values of $q$ and for the  three frequencies given in Table
  \ref{frequencies}. Due to \rf{invrel}, $v_{q,\omega}$ is invariant
  under the interchange $q \leftrightarrow 1/q$.} 
\label{tab:vqomega}
\end{table}

\newpage

\end{document}